# Direct Radiation Pressure Measurements for Lightsail Membranes

Lior Michaeli[1,†], Ramon Gao[1,†], Michael D. Kelzenberg[1], Claudio U. Hail[1], Adrien Merkt[1], John E. Sader[2], and Harry A. Atwater[1,*]

[1]Department of Applied Physics and Materials Science, California Institute of Technology, Pasadena, CA 91125, USA
[2]Graduate Aerospace Laboratories, California Institute of Technology, Pasadena, CA 91125 USA

[†]These authors contributed equally to this work.

*  haa@caltech.edu

## Abstract

Ultrathin lightsails propelled by laser radiation pressure to relativistic speeds are currently the most promising route for flyby-based exoplanet exploration. However, there has been a notable lack of experimental characterization of key parameters essential for lightsail propulsion. Therefore, a model platform for optomechanical characterization of lightsail prototypes made from realistic materials is needed. We propose an approach for simultaneous measurement of optical forces and driving powers, which capitalizes on the multiphysics dynamics induced by the driving laser beam. By modelling the lightsail with a 50-nm thick silicon nitride membrane suspended by compliant micromechanical springs, we quantify force from off-resonantly driven displacement and power from heating-induced mechanical mode softening. This approach allows us to calibrate the measured forces to the driving powers by operating the device as a mechanical bolometer. We report radiation pressure forces of 80 fN using a collimated pump beam of 100 W/cm$^2$ and noise-robust common-path interferometry. As lightsails will inevitably experience non-normal forces, we quantify the effects of incidence angle and spot size on the optical force and explain the nonintuitive trend by edge scattering. Our results provide a framework for comprehensive lightsail characterization and laboratory optomechanical manipulation of macroscopic objects by radiation pressure forces.

## Introduction

Ultrathin lightsails, propelled to relativistic velocities by laser radiation pressure, are being actively explored to enable a new generation of interstellar spacecraft probes, spearheaded by the Breakthrough Starshot Initiative[1,2]. While the achievable speeds of conventional spacecraft technology are limited by ejection of chemical reaction mass, light momentum serves as an alternative, external propellant enabling ultralight spacecraft to reach significantly higher velocities, allowing the reduction of travel time by up to three orders of magnitude for interstellar missions[3]. In contrast to solar sails[4], which rely on radiation pressure from the broadband spectrum of sunlight and its limited irradiance for propulsion, laser-driven lightsails could be accelerated to extreme velocities if propelled by an earth-based kilometer-sized laser array with a power density of $\sim$ MW/cm$^2$, and ultralight weight of a few grams. Therefore, an interstellar lightsail should have a surface area on the order of square meters, which requires extreme width-to-thickness aspect ratios. Consequently, a straightforward realization of lightsails is enabled using subwavelength thick membranes. The Starshot point design case for a flyby mission to our closest known exoplanet[5], Proxima Centauri b, within 20 years of launch demands an infrared laser intensity of MW/cm$^2$ incident on a 10-m$^2$ sized lightsail with a thickness of 100 nm or less.

Previous conceptual and engineering studies have discussed the unprecedented challenges of realizing suitable lightsail membranes from known materials[6–33]. Important criteria include beam-riding stability enabled by structural or photonic engineering[6,7,9–15,17–19,21,25,29,33], thermal management with radiative cooling[8,15,20,22,24,27,30], and sufficient mechanical rigidity[23,28,29]. Therefore, materials for realistic lightsails must balance high reflectance[8,16,24,26,32,33] and low absorption at the propulsion wavelength with low mass density and high tensile strength, while being scalable and compatible with thin-film fabrication. A promising material platform to meet those requirements is silicon nitride, with membranes[34,35] being used extensively for applications in cavity optomechanics[36–38] and nonlinear dynamics[39,40]. The ability to further pattern them with photonic[25,26,41–43] or phononic designs[44–50] or into compliant mechanical resonators[51–55] of up to centimeter in scale[56,57] demonstrates a chip- and wafer-based technological maturity highly desirable for lightsails.

Future lightsail development hinges on quantifying the optomechanical response of lightsail membranes to the propulsive laser beam. Assessing acceleration performance and dynamics requires knowledge of the incident radiation pressure. At the microscopic scale, electrical or optical measurements[58] of radiation pressure forces have been performed in several systems, including cantilevers and mechanical oscillators in ambient environment[59,60], integrated photonic circuits[61], inside an optical cavity[62], for cooling to the quantum ground state of microstructures[63], and to demonstrate force sensitivity below the standard quantum limit[64]. In the context of laser-driven lightsails, radiation pressure forces have been measured on liquid-crystal based, substrate-supported gratings mounted to a torsion oscillator[12,65]. As the optical force is proportional to the incident power, calibration of the measured force to the associated power is needed. This is particularly important when the driving power on the sample changes with the illumination parameters or is not readily measurable, for example, if the size of the laser beam is comparable to the device. Therefore, a comprehensive lightsail characterization platform should enable simultaneous measurement of optical force and power.

The rich history of mechanical manipulation of matter using light, including Arthur Ashkin's optical tweezers[66], enabled trapping and manipulation of microscopic objects of spherical or similar shapes with optical gradient forces originating from a tightly focused laser beam. However, for efficient lightsail propulsion, the radiation pressure interaction needs to be invariant along the axial direction. Additionally, the overlap between beam and lightsail area needs to be optimized over the entire acceleration distance. Such a figure of merit translates to laser light with low angular content, i.e., a nearly collimated beam, with a waist approximately matching the lightsail size. Therefore, optical forces need to be precisely controlled and characterized in this extended optomechanical regime of collimated beams on planar films and nanostructured devices[67–71]. This configuration is naturally scalable from the microscopic to the macroscopic domain, and its exploration is pivotal for advancing the development of dynamical manipulation of millimeter-, centimeter- and ultimately meter-scale objects.

Here, we report quantitative measurements of radiation pressure forces from motion of tens of picometers imparted by a collimated beam impinging on an ultrathin lightsail membrane. The device comprises a square 50-nm-thick silicon nitride membrane, suspended by compliant springs. The sensitive force measurements rely on three key components: rational design of the lightsail as a micromechanical resonator with enhanced mechanical susceptibility, displacement measurements using a noise-robust common-path interferometer with sub-picometer resolution, and an off-resonant driving scheme for exciting quasi-static, linear dynamics. Importantly, our device design enables simultaneous measurement of the driving power based on bolometry via heat-induced mechanical mode softening. We illuminate the lightsail with a continuous-wave (CW) laser power density reaching 100 W/cm$^2$ at a wavelength of 514 nm, and measure the resulting lightsail displacement and optical force. Motion is induced by a collimated laser beam to mimic the conditions under which interstellar lightsails would be accelerated. Furthermore, to better predict the tilt-dependent dynamics of ultrathin lightsails, we characterize the optical force versus incidence angle over the range of ±23°. We show that thin-film interference together with momentum redirection due to edge scattering explains the observed nonintuitive trend in angle dependence of the radiation pressure forces.

## Results

### From interstellar lightsails to laboratory-based lightsails

Chip-scale silicon nitride membranes with thicknesses of 100 nm or less serve as a promising laboratory-based testbed to characterize laser radiation pressure exerted on interstellar lightsails (Fig. 1a). Suspending a square pad representing a microscopic lightsail by tethers offers crucial advantages (Fig. 1b). First, in this architecture, the tethers rather than the membrane deform upon incident radiation pressure. This decouples the effects of optical forces and mechanical bending inherent to membrane dynamics[29], allowing us to focus solely on optical forces in this study (Supplementary Note 1). Second, the pad can move along all translational and rotational degrees of freedom in response to applied forces and torques via bending, stretching and torsion of the tethers. Third, for laser beam sizes comparable to or larger than the pad, edge scattering will occur, whose effect on the radiation pressure force could be characterized. As the laser beam diverges over the acceleration distance, optical scattering from the edges of an interstellar lightsail will be inevitable, making the detection and characterization of edge effects an important aspect of lightsail design. Linearly tethered membranes have been investigated for force sensing and optomechanics applications[51–53]. To quantify radiation pressure forces more precisely, we instead propose incorporating compliant springs, which increase the responsivity of the pad's out-of-plane motion significantly due to enhanced mechanical susceptibility.

### Multiphysics characterization of radiation pressure forces for optomechanics

Our method for simultaneous measurement of optical forces and driving powers capitalizes on the multiphysics dynamics of the mechanical resonator when driven by a laser beam (Fig. 2a). We utilize its dynamical response to the optical drive to quantify force, while its heating-induced thermal response serves as a measure of power. Monitoring the device displacement yields a time trace containing signatures of the off-resonant drive for operation in the quasi-static regime, and of the thermal noise coupling into the device via mechanical modes (Fig. 2b). Spectral analysis of the device displacement reveals features associated with the magnitude of optical force and power (Fig. 2c). Specifically, as the force increases, the amplitude of the power spectral density (PSD) at the drive frequency grows proportionally, while an increase in power results in a red-shift of the resonance frequency via thermal expansion. Our off-resonant excitation also avoids the regime dominated by low-frequency noise while operating above the detection limit (Fig. 2c). Additionally, our excitation scheme minimizes nonlinearities in the mechanical response due to geometrical or material effects of the resonator, parametric excitation due to heat modulation[72–74], and interaction with higher-order mechanical modes[75,76].

We note that the simultaneous measurement of force and power can also be applied in the resonant regime to benefit from the mechanical quality factor, for example, by utilizing two distinct modes to measure force with one and power with the other. The introduced methodology can be employed in macroscopic optomechanics for precise calibration of force measurements as a function of laser beam parameters (Supplementary Note 2).

### Experimental lightsail characterization platform

A robust and sensitive optomechanical, microscope-integrated measurement system is critical for lightsail characterization (Fig. 3). The device under test is a microscopic square lightsail pad (40 μm × 40 μm × 50 nm) suspended by four awl-shaped serpentine springs[77] patterned into a low-stress silicon nitride membrane by electron beam lithography and fluorine-based dry etching (Fig. 3a). Notably, the thickness of 50 nm results in a normal-incidence reflectance of more than 40% for the chosen wavelength, which is close to maximum for unpatterned thin-film silicon nitride (Supplementary Note 3). For detection, we employ a common-path interferometer[78,79], designed to exhibit enhanced immunity to ambient noise (Fig. 3b). A weak unmodulated probe laser beam from a stabilized He-Ne laser (633 nm) is split into two beams using a diffractive beam splitter (DBS), which are subsequently focused onto the microscope image plane, producing two nearly diffraction-limited spots at the sample plane (Fig. 3c). One of these spots is directed onto the pad, while the other is positioned on a stationary reference area of the surrounding rigid silicon substrate. Upon reflection, the sample beam undergoes phase modulation corresponding to the out-of-plane pad motion. The reflected beams are recombined by the same DBS and directed into an avalanche photodiode (APD) to record a dynamic interference signal. The common path of the two beams ensures noise robustness and enables sub-picometer displacement sensitivity. An example of the time-resolved device displacement bandpass-filtered around the off-resonant drive frequency is shown in Fig. 3d, where the quasi-static dynamics is manifested via temporal oscillations of the pad being in phase with the driving force.

To mitigate gas damping, the device is located inside a vacuum chamber of ultra-high vacuum ($5 \times 10^{-9}$ mbar) placed on the linear translation stage of an inverted microscope (Fig. 3e). The microscope interfaces with both excitation and detection beam paths. For excitation, the intensity of a strong CW laser beam at 514 nm is modulated by a feedback-stabilized acousto-optic modulator (AOM). The driving frequency of this modulation is set to $f_d = 4$ kHz, thousands of linewidths (0.3 Hz) below the fundamental mechanical resonance of the device ($f_m = 9.4$ kHz). The polarization state and intensity are controlled by a half-wave plate followed by a polarizer. The excitation beam is focused onto the objective lens's back-focal plane to generate a collimated beam for driving the device. The voltage readout was converted to units of displacement according to the range of $\lambda/4$ for double-sided phase wrapping (see Methods).

## Measurement of optical forces by simultaneous power calibration

The force $F$ exerted by a plane wave incident upon an object that transmits, absorbs, and reflects light specularly depends on the beam's power $P$ and the object's optical properties. Specifically, the propulsive radiation pressure force on the lightsail is given by[80]:

$$F_z(\theta) = \frac{P(\theta)}{c}[2R(\theta) + A(\theta)]\cos\theta, \qquad (1)$$

where $R$ and $A$ are the angle-dependent reflection and absorption, respectively, $c$ is the speed of light, and the cosine term accounts for the projection of the force along the $z$ axis (Fig. 1a). The force in our case is governed by the reflection (~40%), which is orders of magnitude larger than the absorption (Supplementary Note 3). We note that recoil radiation pressure from thermal emission is also negligible due to the low absorption, small temperature increase and symmetric emissivity of both sides of the membrane (Supplementary Note 3). We measure the lightsail pad displacement $z$ upon illumination and find the radiation pressure force according to $F_z = k_m \left(1 - \left(\frac{f_d}{f_m}\right)^2\right) z$. The stiffness (spring constant) of the membrane $k_m$ is extracted from the stochastic thermal noise excited fundamental resonance using the equipartition theorem[81]. The measured displacement and extracted force increase linearly with driving laser power (Fig. 4a), confirming operation in the linear regime. To characterize the force versus lightsail-tilt-angle $\theta$, the driving power should be kept constant. However, in experiment, the power inevitably changes with incidence angle. Moreover, we aim to study a range of cases, from the beam spot being fully contained within the pad to the beam spot overfilling the pad. To precisely determine the driving power experienced by the pad, we simultaneously use the device as a mechanical bolometer[82–84] (Fig. 2). Specifically, the Lorentzian-like resonance peak shifts to lower frequencies for increasing drive power due to absorption (Fig. 4b). The resonance frequency varies linearly with laser power according to a responsivity of $\mathcal{R} = -0.9$ Hz/μW (Fig. 4c). Measurement of optical forces for varying illumination parameters by simultaneous power calibration is further detailed in Supplementary Note 2.

Characterization of angle-dependent radiation pressure forces and edge scattering

Interstellar lightsails will experience non-normal forces due to perturbations and beam-lightsail misalignment over the course of their acceleration. The direction and magnitude of these forces determine the trajectory, with inclusion of photonic designs enabling beam-riding stability. To generate oblique illumination conditions, the laser spot is moved in the objective's back-focal plane (Fig. 3e). This allows setting the incidence angle $\theta$ of the beam on the pad within the numerical aperture of the objective (±24°). The Fourier plane of the image confirms both the incidence angle and the collimated nature of the beam with an angular spread of $\Delta\theta = 1.5°$ (Fig. 5a). In real space, we control the beam spot size to underfill the pad area at normal incidence.

Using this approach, we characterize the radiation pressure force on the tethered lightsail as a function of incidence angle $\theta$ (Fig. 5b top, see Supplementary Notes 4 and 5 for more details). For constant laser power on the pad, the angle-dependent force projected along the $z$-axis follows a cosine relationship (gray curve), see equation (1). Due to the lightsail's subwavelength thickness, thin-film interference alters the angle-dependent reflection (Fig. 5b bottom, Supplementary Note 3), further modifying the force. Nonetheless, our measurements show a much steeper fall-off with angle. To investigate this discrepancy at larger incidence angles, we image the laser spot sizes with increasing incidence angle. Figure 5c shows the laser beam overfilling the pad aperture at angles exceeding 16°, leading to significant scattering of light from the pad edges. This beam expansion is due to the oblique beam being projected on the pad and the change of focal plane when illuminating through the vacuum chamber window. The light scattering from the edges of the pad contributes to momentum transfer in directions different from those of transmission and specular reflection. To understand the effect of edge scattering on the out-of-plane force, we performed full-wave electromagnetic simulations in COMSOL Multiphysics based on the experimentally measured laser beam sizes (Supplementary Note 6). The shaded area corresponds to an uncertainty of 12 μm in centering the beam on the pad due to optical aberrations when imaging through the vacuum chamber's window. To further confirm the effect of edge scattering on the radiation pressure, we report measurements at 23° on lightsails with varying pad sizes of 100-nm thickness (Fig. 5d, see Supplementary Note 7 for optical images). As the pad size increases, the measured force approaches the theoretical force for an infinitely extended planar film (blue line). The observed significant modification of the radiation pressure force in the acceleration direction (normal to the pad) due to edge scattering highlights the importance of our model platform to characterize such effects in the early stages of lightsail development and prototyping.

## Discussion

We present a direct characterization of radiation pressure forces on silicon nitride lightsail membranes, using the device itself to simultaneously measure the incident power. Specifically, we propose membranes suspended by compliant springs as a susceptible and multiphysics testbed for precise measurements of radiation pressure forces and associated power. Our fabricated device represents a scalable chip-based design enabling comprehensive studies of the light-matter interaction within a lightsail from the microscopic to the macroscopic scale. Displacements of up to 50 pm are resolved for incident radiation pressure of up to 100 W/cm² using a common-path interferometer designed for robustness to mechanical noise and sensitivity to weak forces. By driving the device sufficiently below the fundamental mode, we operate in the quasi-static linear regime necessary for modelling radiation pressure forces on realistic lightsails. Moreover, we demonstrate control of various properties of the excitation laser beam, including its incidence angle, collimation, and spot size as a toolbox for comprehensive characterization of lightsail dynamics.

By quantifying the device stiffness from the equilibrium stochastic fluctuations of the fundamental mode, we determine radiation pressure forces of up to 80 fN. Linear dependence of the optical force on the applied laser power is accompanied by a linearly decreasing resonance frequency, demonstrating the simultaneous operation of our lightsail prototype as a sensitive micromechanical bolometer with a responsivity of $\mathcal{R} = -0.9$ Hz/μW. The power-reporting mechanism of our device is crucial for correctly normalizing the input power and precisely quantifying the radiation pressure force versus incidence angle. As illumination becomes more oblique, the initially underfilling excitation laser beam starts to overfill the lightsail pad, resulting in edge scattering, which significantly alters the radiation pressure force. By quantifying this unanticipated trend of angle-dependent radiation pressure for different beam sizes, we demonstrate the importance of accurately characterizing forces on candidate materials for lightsail applications.

Our observation platform enables characterization of the mechanical, optical, and thermal properties of lightsail prototype devices, thus opening the door for further multiphysics studies of radiation pressure forces on macroscopic objects. Additionally, photonic, phononic or thermal designs tailored to optimize different aspects of lightsailing can be incorporated and characterized. In particular, characterizing and shaping optical forces with nanophotonic structures for far-field mechanical manipulation is central to the emerging field of meta-optomechanics, allowing for arbitrary trajectory control of complex geometries and morphologies with light[67–70]. Laser-driven lightsails require self-stabilizing forces and torques emerging from judiciously designed metasurfaces for beam-riding. We expect that their direct observation is possible using our testbed, which is an important steppingstone towards the realization of stable, beam-riding interstellar lightsails, and optomechanical manipulation of macroscopic metaobjects.

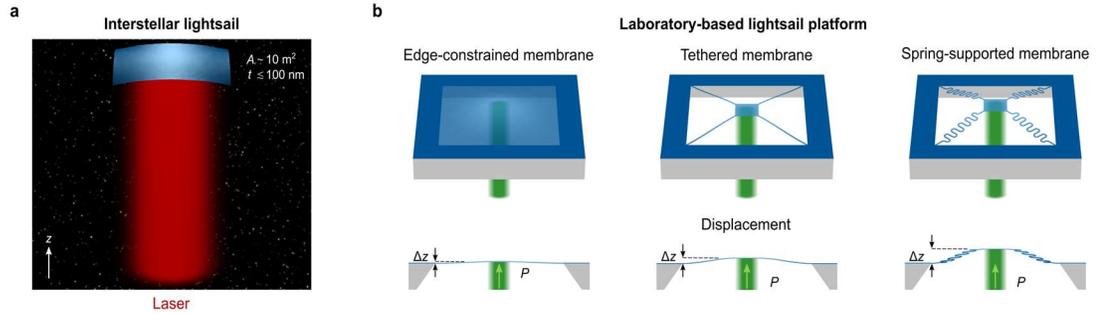

**Fig. 1: From interstellar lightsails to laboratory-based lightsail platforms. a**, Concept of laser-propelled interstellar lightsail of 10 m$^2$ in area and 100 nm or less in thickness. **b**, Laboratory-based lightsail platforms relying on edge-constrained silicon nitride membranes (left), linearly tethered membranes (middle) and spring-supported membranes (right). Removing the edge constraint allows to decouple the effects of optical force and membrane deformation, model lightsail dynamics, and study optical scattering from the edges. Suspending lightsails by compliant serpentine springs rather than linear tethers significantly increases its mechanical susceptibility to laser radiation pressure of the same power $P$, resulting in larger out-of-plane displacement $\Delta z$ for more precise detection.

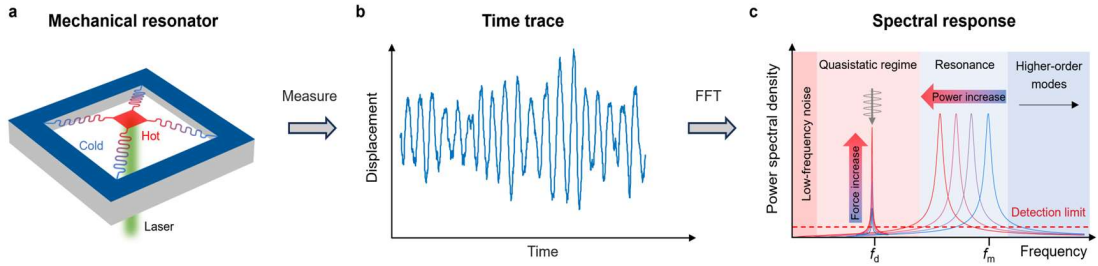

**Fig. 2: Multiphysics platform for radiation pressure characterization in optomechanics. a**, A susceptible mechanical resonator displaces and heats up in response to an incident laser beam. **b**, The displacement time trace of the resonator exhibits signatures of the drive modulation frequency and the thermal noise. **c**, The power spectral density of the displacement reveals distinguishable contribution from the applied optical force and optical power. The force magnitude is manifested in the amplitude of the spectral peak at the drive frequency, while the power magnitude determines the shift of the resonance frequency, enabled by heat-induced mode softening. An off-resonant excitation scheme is employed to operate in the quasi-static regime, which also enhances detection accuracy by circumventing nonlinearities inherent to resonant excitation and suppressing interaction with higher-order modes.

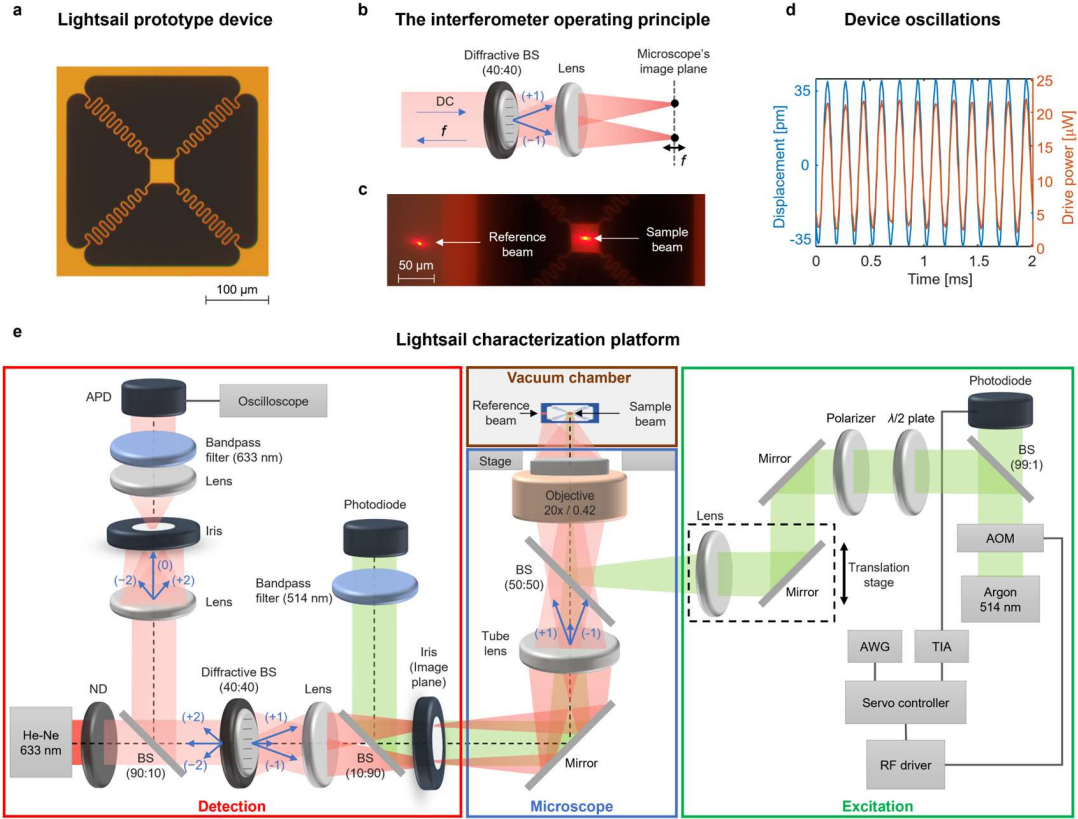

**Fig. 3: Lightsail prototype device and experimental characterization platform. a**, Microscope image of the fabricated 50-nm-thick lightsail prototype device based on a spring-supported silicon nitride membrane. **b**, The operating principle of the common-path interferometer with high immunity to ambient noise. The unmodulated (DC) probe laser beam is split into two beams (−1 and +1) using a diffractive beam splitter (DBS) and subsequently focused onto the microscope's image plane. This results in two closely spaced, nearly diffraction-limited spots, with one positioned on the pad and the other on a reference area (silicon nitride on silicon substrate). The reflected laser beam undergoes phase modulation corresponding to the pad's out-of-plane motion at frequency $f$. **c**, Microscope image of the detection and reference beams on the sample plane. **d**, Measured time-resolved pad displacement bandpass-filtered around the drive frequency. The motion is in phase with the incident drive signal, confirming operation in the quasi-static regime. **e**, The experimental setup comprises three parts: an excitation path with a pump laser beam from an Ar-ion laser (514 nm) modulated in intensity using feedback-controlled acousto-optic modulator (AOM), a detection path based on a common-path interferometer with a probe laser beam from a stabilized He-Ne laser (633 nm), and a vacuum stage with ultra-high vacuum ($5\times10^{-9}$ mbar) on an inverted microscope platform containing the lightsail prototype. The drive frequency is set to 4 kHz, below the fundamental resonance of the lightsail at 9.4 kHz. Collimation of the excitation beam on the pad is achieved by focusing the beam onto the back-focal plane of the objective, with its incidence angle controlled by a linear translation stage.

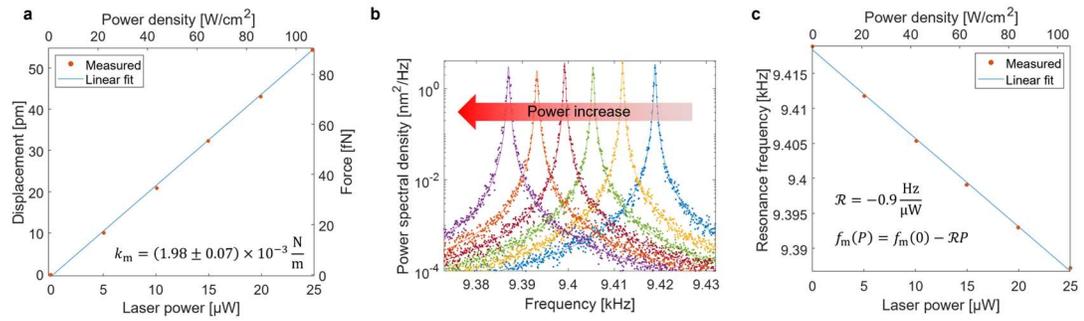

**Fig. 4: Measurement of optical force on power-reporting lightsails. a**, Lightsail pad displacement versus laser power (bottom axis) or power density (top axis) showing a linear relationship. Displacement is converted to force according to the stiffness $k_m$ of the tethered device extracted from the thermal noise power spectral density and weighted by the off-resonant driving frequency. **b**, The fundamental resonance peak shown by the power spectral density and excited by thermal noise is shifted to lower frequencies with increasing laser power due to heat-induced mechanical softening. **c**, An observed linear decrease of resonance frequency $f_m$ versus laser power $P$ or power density, validates operation of the lightsail pad as a micromechanical bolometer with a responsivity of $\mathcal{R} = -0.9$ Hz/µW. The linear fits in **a** and **c** were obtained by least-squares minimization.

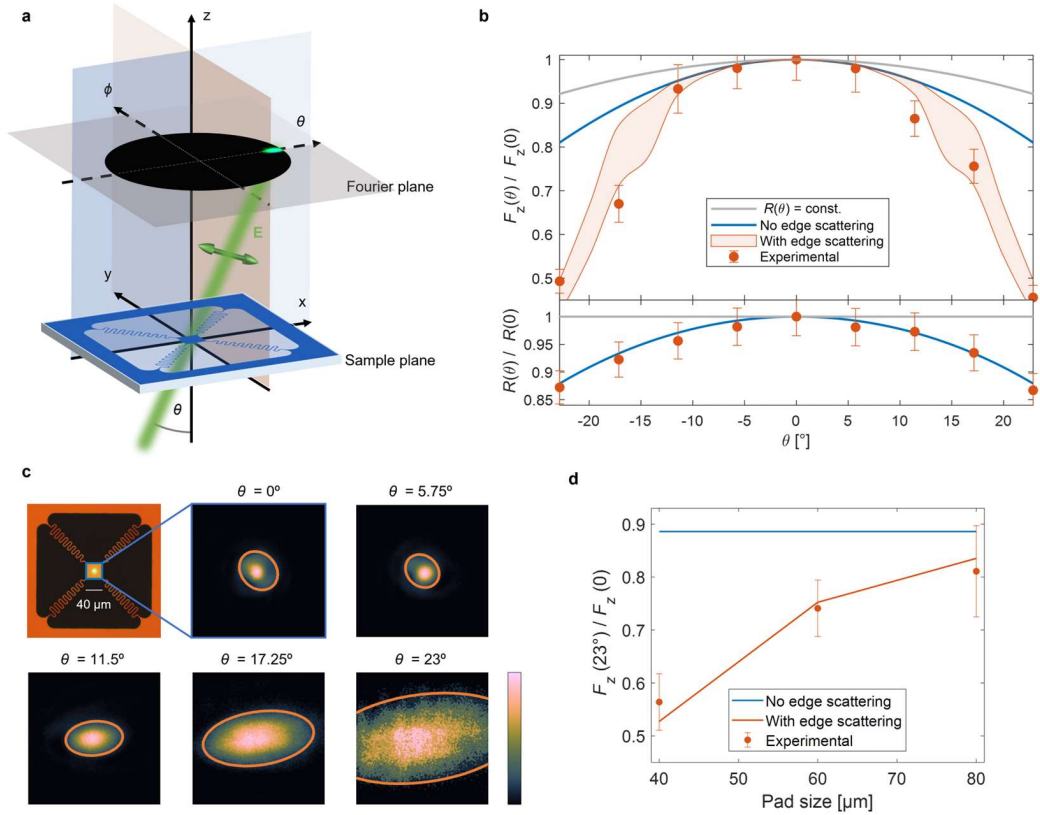

**Fig. 5: Characterization of angle-dependent radiation pressure forces for lightsail propulsion**. **a**, Schematic of the oblique laser beam (transverse-magnetic polarization according to orientation of the electric field **E**) incident on tethered lightsail, with the Fourier plane revealing the incidence angle $\theta$ and angular content of the beam $\Delta\theta$. **b**, Out-of-plane force $F_z$ (top) and reflectance $R$ (bottom) as a function of the angle of incidence $\theta$. Force and reflectance are normalized with respect to normal incidence values. The error bars represent the propagated standard error (Supplementary Note 5). Deviation of measured force (orange points) from expected trend (blue curve) despite measured reflectance following theory can be explained by edge scattering as simulated in COMSOL based on experimentally determined beam sizes. **c**, Experimentally measured beam size versus angle of incidence, with orange contour lines indicating ellipsoidal fit of laser spot size. The images were taken on the substrate, whereas potential beam-pad misalignment of 12 μm was accounted for to calculate the shaded area in **b**. Each image is normalized to its own maximum. **d**, Out-of-plane force at 23° normalized to force at normal incidence as a function of the lightsail pad size. The impact of edge scattering is reduced as the pad size increases for a constant beam size.

# Methods

### Device fabrication

A 200-nm thick, positive-tone resist (ZEP 520A, D.R. 1.5) is spun onto low-stress silicon nitride membranes of 100 nm thickness (Norcada, NXA10040C), followed by 200 nm of aquaSAVE to prevent charging during electron beam lithography (280 μC/cm2). After lithography, the conductive polymer is removed by dipping the device in DI water, followed by development in ZED-N50 for two minutes and 30 seconds and subsequent cleaning in IPA. The lithographically defined pattern is then transferred into the membrane via ICP-RIE (pseudo-Bosch etch for 2:30 min in SF6/C4F8 chemistry). To prevent contamination of the membrane's backside with the thermal oil (Santovac), the membrane is glued onto two silicon stilts. Finally, the resist is stripped via $O_2$ plasma etching for 90 minutes. The final thickness of the membrane is 50 nm given the etching conditions.

### Experimental setup

All measurements are conducted using an automated experimental setup. Specifically, the mixed-domain oscilloscope (Tektronix MDO32), the arbitrary waveform generator (Keysight, 33220A AWG) and the motorized rotational stage of the half-wave plate mount (Thorlabs, PRM1Z8) are controlled with dedicated MATLAB scripts.

### Detection path: displacement measurement using common-path interferometry

To detect lightsail displacement, free space common path interferometer is employed. To generate the sample and reference probe beams, light from a stabilized HeNe (Thorlabs HRS015B, 633 nm, polarized, 1.2 mW) is split using a binary phase grating DBS (Holor/Or, fused silica, anti-reflection coated) and coupled into the microscope (Zeiss, Axio Observer Z1). The laser attenuation is set using neutral density filters to probe the pad motion with a power of ~ 1 μW. The reflected HeNe beams are detected by a temperature-compensated silicon avalanche photodiode (Thorlabs APD410A) and recorded by the mixed-domain oscilloscope. The HeNe path between the DBS and the microscope is enclosed with cardboard to reduce perturbations to the interferometer due to air flow and temperature gradients.

### Detection path: reflectance measurement

The power reflected from the sample was measured with a silicon photodiode (Thorlabs, SM05PD1A) connected to a transimpedance amplifier (Thorlabs, PDA200C).

### Excitation path

The device is driven by an CW argon-ion laser at 514 nm (Innova 70c) intensity-modulated by an AOM (NEOS 23080-1) connected to a feedback-controlled RF Driver (Isomet, 232A-2). The feedback is configured with a closed loop (Fig. 3e) using a servo controller (Newport, MKS, LB1005) for monitoring the power of the modulated beam with a silicon photodiode (ams-OSRAM, BPW 34 S-Z, soldered in parallel to a 1 kΩ resistor) connected to a transimpedance amplifier (TIA) (Edmunds Optics, SN 57-601), and for correcting accordingly the input signal to the AOM to match the desired waveform set by the arbitrary waveform generator (AWG).

## Conversion of displacement from voltage to length units

To convert the displacement data recorded by the oscilloscope in voltage units to length units, we rely on the $\lambda/4$ phase range of the interferometer signal being single valued, as described in the following. The phase of the reflected sample and reference beams is determined according to the pad's motion:

$$\phi(t) = 2kz(t), \tag{2}$$

where $k = 2\pi/\lambda$ is the wavevector and $\lambda$ is the wavelength. Thus,

$$z(t) = \frac{\lambda}{4\pi}\phi(t). \tag{3}$$

The intensity recorded at the detector follows:

$$I(t) = I_R + I_S + 2\sqrt{I_R I_S} \cdot \cos(\phi_0 + \phi(t)), \tag{4}$$

where $I_R$ and $I_S$ are the static intensities of the reference and detection beams, respectively, and $\phi_0$ is a phase offset working point, determined according to the optical alignment. The working point is set to $\pm \pi/2$ for maximum sensitivity. The light falling on the APD is detected and amplified, such that a voltage linearly proportional to the light's intensity is recorded by the oscilloscope:

$$v(t) = \eta I(t), \tag{5}$$

where $\eta$ is a conversion factor, accounting for the APD detection sensitivity and amplification. To extract $\phi(t)$ from the measured signal, it is convenient to cast equations (4) and (5) into the following form:

$$v(t) = A\cos(\phi_0 + \phi(t)) + B, \tag{6}$$

where $A = 2\eta\sqrt{I_R I_S}$ and $B = \eta(I_R + I_S)$. To find experimentally the value of these constants, another measurement is performed with the driving laser modulated at the resonance frequency. Due to its high susceptibility, the pad oscillations are large, probing the full $\cos(\phi_0 + \phi(t)) \in [-1,1]$ range. We denote this measurement as $v_{\text{folded}}(t)$. The lower and upper range of the recoded values correspond to $\cos(\phi_{\min}) = -1$ and $\cos(\phi_{\max}) = 1$, correspondingly:

$$a \equiv \min(v_{\text{folded}}(t)) = -A + B, \quad b \equiv \max(v_{\text{folded}}(t)) = A + B. \tag{7}$$

By solving for $A$ and $B$, we find:

$$A = \frac{b-a}{2}, \quad B = \frac{a+b}{2}. \tag{8}$$

Finally, we extract $\phi(t)$ from equations (6) and (8), and substitute the result into equation (3) to obtain:

$$z(t) = \frac{\lambda}{4\pi}\left(\cos^{-1}\left(\frac{v(t)-B}{A}\right) - \phi_0\right). \tag{9}$$

## Time series acquisition and spectral analysis

The time series of the off-resonant drive (4 kHz) consists of $10^7$ data points, sampled at a frequency of 100 kHz, and is captured using the oscilloscope's high-resolution mode. To extract the full phase folding range, another measurement with the drive set to the resonance frequency is performed. This measurement consists of $10^6$ data points, with the other parameters unchanged.

To determine the membrane's displacement and its resonance frequency, we analyze the signal in the spectral domain. The displacement is extracted from the amplitude at 4 kHz and the resonance frequency from the spectral position of the corresponding Lorentzian-like peak.

## Device stiffness measurement

Thirty 100-second-long time-series measurements of the thermal noise fluctuations of the pad are acquired, from which its power spectral density is determined. This time series is sampled at 100 kHz. The fundamental mode resonance peak is fitted to a Lorentzian line-shape, and the area under the curve specifies the variance of the mechanical displacement $\langle z^2 \rangle$. The device stiffness is then calculated according to the equipartition theorem by $k_\mathrm{m} = \frac{k_\mathrm{B} T}{\langle z^2 \rangle}$, where $k_\mathrm{B}$ is Boltzmann's constant and $T$ is temperature. The extracted value is $k_\mathrm{m} = (1.98 \pm 0.07) \times 10^{-3}$ N/m.

The stiffness also satisfies $k_\mathrm{m} = m \omega_\mathrm{m}^2$, where $m$ is the effective mass and $\omega_0 = 2\pi f_\mathrm{m}$ is the angular resonance frequency. Therefore, the relative change in spring constant can be estimated by $\Delta k_\mathrm{m}/k_\mathrm{m} = 2\Delta f_\mathrm{m}/f_\mathrm{m}$. For a resonance frequency of $f_\mathrm{m} = 9.4$ kHz with a maximal change of $\Delta f_\mathrm{m} = 30$ Hz, a change of 0.64% in $k_\mathrm{m}$ is expected. This value is smaller than the uncertainty of 3.53% in our stiffness determination. The spring constant is constantly monitored during all measurement to ensure a constant value for different driving laser properties, i.e., power, incidence angle and spot size.

## Modelling of edge scattering

The effect of edge scattering on the normal radiation pressure force is simulated in COMSOL Multiphysics. The background field formulation in the Wave Optics module is employed and a Gaussian beam is defined according to the experimentally measured beam size versus incidence angle. The 50-nm thick pad is placed at the center of a circular simulation domain surrounded by perfectly matched layers and rotated according to the incidence angle. Closed circular paths are set up around the pad, over which the Maxwell stress tensor is integrated to obtain the radiation pressure. The optical force is calculated by integrating the product of radiation pressure and laser beam intensity over the area of the pad.

Perceptually uniform color maps were used for Fig. 5[85].

## Data availability

The data supporting the findings of this study are available from the corresponding author upon request.

## Acknowledgements

This work was supported by the Air Force Office of Scientific Research under grant FA2386-18-1-4095 and the Breakthrough Starshot Initiative.

LM acknowledges support of the Fulbright Israel Postdoctoral Fellowship. The authors acknowledge helpful discussions with Keith Schwab, Marianne Aellen, Cora Wyent, Yonghwi Kim and Avi Michaeli.

## Author information

### Authors and Affiliations

**Department of Applied Physics and Materials Science, California Institute of Technology, Pasadena, CA 91125, USA**
Lior Michaeli, Ramon Gao, Michael D. Kelzenberg, Claudio U. Hail, Adrien Merkt & Harry A. Atwater

**Graduate Aerospace Laboratories, California Institute of Technology, Pasadena, CA 91125 USA**
John E. Sader

### Contributions

L.M., R.G. and H.A.A. conceived the project. L.M. built the common-path interferometer. R.G. fabricated the devices. L.M. and R.G. built the optical setup, conducted the measurements, and performed the data analysis. M.D.K. assisted with the feedback stabilization of the acousto-optic modulator. M.D.K. and C.U.H. provided inputs to the optical setup and measurements. M.D.K. and A.M. designed the vacuum chamber and line. J.E.S. offered guidance with the device stiffness measurement and provided theoretical insights. L.M. and R.G. wrote the manuscript with inputs from all authors. H.A.A. supervised the project.

### Corresponding author

Correspondence to Harry A. Atwater.

## Ethics declarations

### Competing interests

The authors declare no competing interests.

## Supplementary information

File is attached.

Supplementary Notes 1-7, Supplementary Figures 1-6, and Supplementary references 1-6.

# Supplementary Information: Direct Radiation Pressure Measurements for Lightsail Membranes

Lior Michaeli[1,†], Ramon Gao[1,†], Michael D. Kelzenberg[1], Claudio U. Hail[1], Adrien Merkt[1], John E. Sader[2], and Harry A. Atwater[1,*]

[1]Department of Applied Physics and Materials Science, California Institute of Technology, Pasadena, CA 91125, USA
[2]Graduate Aerospace Laboratories, California Institute of Technology, Pasadena, CA 91125 USA

[†]These authors contributed equally to this work.

* haa@caltech.edu

## Supplementary Note 1: COMSOL simulation of static membrane deformation

COMSOL Multiphysics simulations are performed to establish negligible pad deformation relative to the springs. Specifically, we employ a coupled mechanical-thermal prestressed stationary study, whereby we (i) relax the uniform stress along the springs, then (ii) apply a force corresponding to the incident laser in a second step, and (iii) simulate the relative displacement of the pad for illumination conditions and maximum power according to Fig. 4. The calculated pad displacement (relative to its maximal value) is given in Fig. S1. This reveals negligible deformations with a maximum relative displacement across the pad of ~ 1% from center to edge.

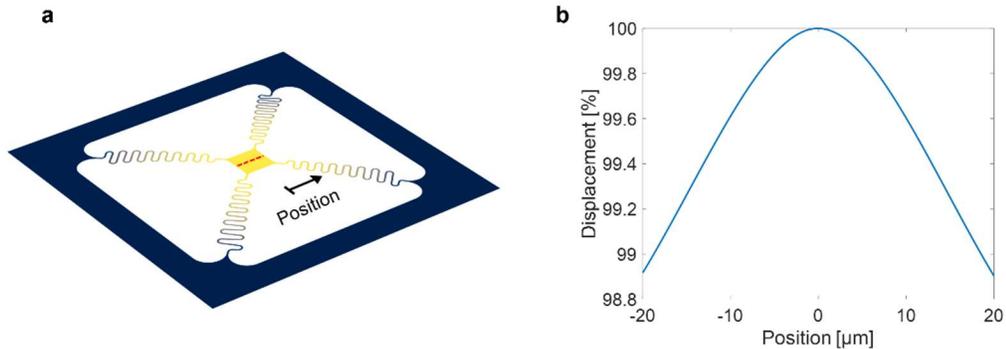

**Fig. S1 Simulated relative displacement within the pad upon incident radiation pressure force. a**, Visualization of the device's color-coded displacement, showing nearly uniform displacement across the pad. **b**, relative displacement as a function of position on the pad, evaluated along the red dashed line depicted in **a**. The simulations reveal negligible deformations of the pad, with displacement variations of ~ 1% from center to edge, relative to the frame.

# Supplementary Note 2: Measurement of optical forces for varying illumination parameters by simultaneous power calibration

Here we describe our methodology to measure the optical force $F$ applied to the pad as a function of the illumination parameters, e.g., angle of incidence, angular content, polarization state and spot size. We cast all parameters into a single parameter vector, denoted as $\boldsymbol{S}$. To characterize $F(\boldsymbol{S})$, one needs to perform force measurements while maintaining the driving laser power constant. Particularly, if only a portion of the beam illuminates the membrane, it is important to maintain a constant power for this fraction as a function of the parameters, i.e., $P(\boldsymbol{S}) = P_0$. This enables determination of $F(P_0, \boldsymbol{S})$, i.e., the force dependence on the parameters $\boldsymbol{S}$ for a constant power. However, in practical experimental realizations, the driving laser power inevitably changes with the driving beam parameters such that only $F(P(\boldsymbol{S}), \boldsymbol{S})$ can be directly measured. This poses a challenge to isolate the direct dependence of the optical force on the illumination parameters $\boldsymbol{S}$. In the following, we describe our approach to achieve this goal.

We rely on the theoretical observation that for constant (fixed) material properties, the optically induced force depends linearly on the driving optical power:

$$F = S_{F-P} P, \qquad (S1)$$

where the force-to-power slope is $S_{F-P}$. This is a direct consequence of the momentum conservation of matter and electromagnetic fields[6], as also evident from equation (1) in the main text. The experiments are designed to validate the linearity for each measurement. Thus, according to equation (S1), it is sufficient to measure the driving power $P$ to fully calibrate the force to a constant power of interest $P_0$ and obtain the direct dependence on the parameters $\boldsymbol{S}$:

$$F(P_0, \vec{S}) = F(P(\boldsymbol{S}), \boldsymbol{S}) \frac{P_0}{P(\boldsymbol{S})}. \qquad (S2)$$

Therefore, our primary challenge lies in accurately measuring the driving power $P(\boldsymbol{S})$. To address this challenge, we utilize the membrane as a micromechanical bolometer where its fundamental mechanical resonance frequency $f_0$ changes in response to variations in the driving optical power $P$. This approach allows us to collect force and power data simultaneously from the same time signal, each corresponding to different spectral locations. To elaborate, force data is extracted from the off-resonant drive frequency, while power data is measured by monitoring the fundamental resonance frequency. To leading order, the change in the fundamental resonance frequency $f_0$ as a function of the optical power $P$ is given by:

$$f_0(P) = f_0(0) - \mathcal{R} P, \qquad (S3)$$

where $\mathcal{R}$ is the responsivity of the resonance frequency to the driving power. Once $\mathcal{R}$ is determined for a configuration where $P$ is known, equation (S3) can be utilized to calculate the power for other configurations. In our experiment, the power is accurately measured with an external power meter (Thorlabs, PM16-120) under condition of normal incidence illumination with the entire laser beam spot being contained within the membrane. While this outlines the main elements of our methodology, it is essential to consider certain practical aspects, as further elaborated below.

The linearity of equations (S1) and (S3) is central to our methodology. While equation (S1) is theoretically valid, experimental validation is required to ensure its accuracy in the presence of potential noise sources and measurement inaccuracies. Moreover, the linearity of equation (S3) holds true only for sufficiently small drive powers, necessitating verification of operation within the linear regime.

Therefore, we conduct measurements of force and resonance frequency versus the driving power for each experimental setting.

For each configuration of the parameters $\boldsymbol{S}$, the power is some value $P_0(\boldsymbol{S})$, which is determined by the optical alignment. We take advantage of our ability to alter the driving power without affecting the optical path. For example, we control the power by rotating the motorized half waveplate mount prior to a fixed polarizer, such that the undetermined power $P_0(\boldsymbol{S})$ satisfies the relation:

$$P(\boldsymbol{S}, \alpha) = P_0(\boldsymbol{S}) \cdot \alpha, \tag{S4}$$

where $\alpha \in [0,1]$ is a power control parameter. Combining equations (S1) and (S4), gives:

$$F(\boldsymbol{S}, \alpha) = S_{F-P} P_0(\boldsymbol{S}) \cdot \alpha. \tag{S5}$$

In addition, by substituting equation (S4) to equation (S3) we obtain:

$$f_0(P(\boldsymbol{S}, \alpha)) = f_0(0) - \mathcal{R} P_0(\boldsymbol{S}) \cdot \alpha. \tag{S6}$$

Thus, by scanning $\alpha$ in the range from 0 to 1, we are able to verify the linear relationships in equations (S1) and (S3) and measure the corresponding slopes $S_{F-\alpha}(\boldsymbol{S}) = S_{F-P} P_0(\boldsymbol{S})$ and $S_{f_0-\alpha}(\boldsymbol{S}) = -\mathcal{R} P_0(\boldsymbol{S})$. Taking their ratio removes $P_0(\boldsymbol{S})$:

$$S_{F-P}(\boldsymbol{S}) = -\frac{S_{F-\alpha}(\boldsymbol{S})}{S_{f_0-\alpha}(\boldsymbol{S})} \mathcal{R}. \tag{S7}$$

Up to this point, a constant responsivity $\mathcal{R}$ is assumed. However, the responsivity depends on the pad's morphology and absorption, with the latter generally depending on $\vec{S}$. For precise measurement, this dependence should be quantified. In our experiments, this dependence is weak (< 6%), but is still considered (see Supplementary Note 4). Importantly, once the trend of $\mathcal{R}$ versus $\boldsymbol{S}$ is measured, one needs to extract $\mathcal{R}$ only for one known configuration in which the power can be determined precisely with a standard power meter.

## Supplementary Note 3: Optical properties of thin-film silicon nitride membranes

### Theoretical reflectance and absorption from optical transfer matrix method

The theoretical reflectance $R$ and absorption $A$ of thin-film silicon nitride membranes are calculated using the optical transfer matrix method[1] (TMM) and given by

$$R = |r|^2, \quad r = \frac{(M_{11} + M_{12}\eta_3)\eta_1 - (M_{21} + M_{22}\eta_3)}{(M_{11} + M_{12}\eta_3)\eta_1 + (M_{21} + M_{22}\eta_3)}, \tag{S8}$$

$$A = \frac{4\eta_1 \operatorname{Re}[(M_{11} + M_{12}\eta_3)(M_{21} + M_{22}\eta_3)^* - \eta_3]}{|\eta_1(M_{11} + M_{12}\eta_3)(M_{21} + M_{22}\eta_3)|^2}, \tag{S9}$$

where **M** is the optical transfer matrix for a slab suspended in air, given by

$$\mathbf{M} = \begin{pmatrix} M_{11} & M_{12} \\ M_{21} & M_{22} \end{pmatrix} = \begin{pmatrix} \cos(\delta) & \dfrac{i}{\eta_2}\sin(\delta) \\ i\eta_2 \sin(\delta) & \cos(\delta) \end{pmatrix}, \quad (S10)$$

with

$$\boldsymbol{\eta} = (\eta_1, \eta_2, \eta_3)^T = \frac{1}{Z_0}\left(\frac{1}{\sqrt{1-\sin^2(\theta)}}, \frac{\tilde{n}}{\sqrt{1-(\sin(\theta)/\tilde{n})^2}}, \frac{1}{\sqrt{1-\sin^2(\theta)}}\right)^T, \quad (S11)$$

and the phase factor $\delta = -\frac{2\pi}{\lambda}\tilde{n}d\sqrt{1-(\sin(\theta)/\tilde{n})^2}$, with $\lambda$ being the wavelength, $\tilde{n} = n - ik$ the complex refractive index, $d$ the thickness of the membrane and $\theta$ the incidence angle of the beam.

Theoretical dependence of reflectance on membrane thickness

The reflectance of the silicon nitride membrane is governed by thin-film interference within the layer, as shown in Fig. S2. The calculation is performed using the TMM, assuming a complex refractive index according to the ellipsometry measurements ($n = 2.22$ and $\kappa = 3.5 \times 10^{-3}$). Multiple peaks of the reflectance as a function of the layer thickness are seen, corresponding to different orders of interference within the layer. In our experiment, we operate at a thickness of 50 nm, in vicinity to the first peak at $\lambda/(4n)$ (black vertical line), to maximize the reflectance while ensuring minimal lightsail mass.

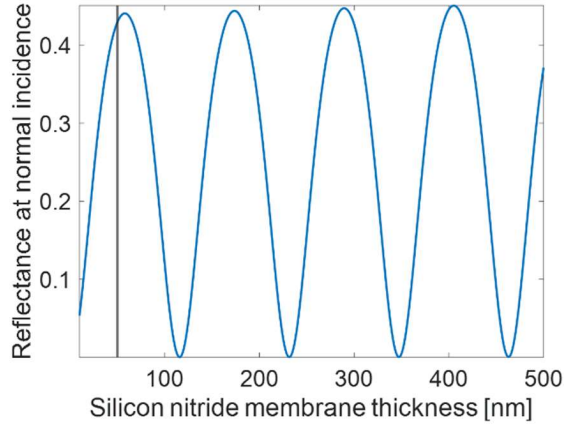

**Fig. S2 Theoretical dependence of reflectance on membrane thickness.** Thin-film interference within the layer produce periodic variation of the reflectance. The experiments are performed for a membrane thickness of 50 nm (black vertical line) to maximize reflectance while keeping a minimal mass.

Measurement of angle-dependent reflectance

The reflection of the pad versus the angle of incidence is shown in Fig. S3. The measurements are performed using:

$$R(\theta) = \langle R_{\text{sam}}^{\text{exp}}(\theta)/R_{\text{sub}}^{\text{exp}}(\theta)\rangle R_{\text{sub}}^{\text{theo}}(\theta), \tag{S12}$$

where $R_{\text{sam}}^{\text{exp}}$ and $R_{\text{sub}}^{\text{exp}}$ are the measured reflected power of the pad and substrate, correspondingly, and $R_{\text{sub}}^{\text{theo}}$ is the theoretical reflection of the substrate. The substrate is a 100-nm thick silicon nitride layer on top of a thick silicon layer. We perform three ten-second-long measurements (5 million samples) of reflected power from the pad and substrate per incidence angle upon incident illumination modulated at $f_d$ = 4 kHz, at ambient pressure without the vacuum chamber glass. Then, the short-time Fourier transform is calculated for each sample (pad) and reference (substrate) signal by dividing them into Kaiser windows of 1 ms, from each of which the amplitude at 4 kHz was extracted. Doing so allow us to obtain a time-resolved relative reflectance, $R_{\text{sam}}^{\text{exp}}(\theta)/R_{\text{sub}}^{\text{exp}}(\theta)$, which is subsequently averaged to obtain the reflectance for a specific incidence angle.

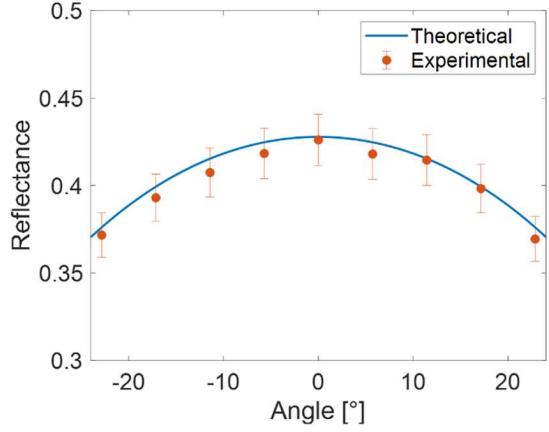

**Fig. S3 Experimental angle-dependent reflectance.** Measured reflectance versus incidence angle on top of TMM calculation.

Independence of normalized angle-resolved absorption on extinction coefficient

The membrane's angle-dependent absorption normalized to the absorption at normal incidence for varying extinction coefficients are shown in Fig. S4. The calculations are performed using the TMM with a layer thickness of 50 nm and a refractive index of $n$ = 2.22.

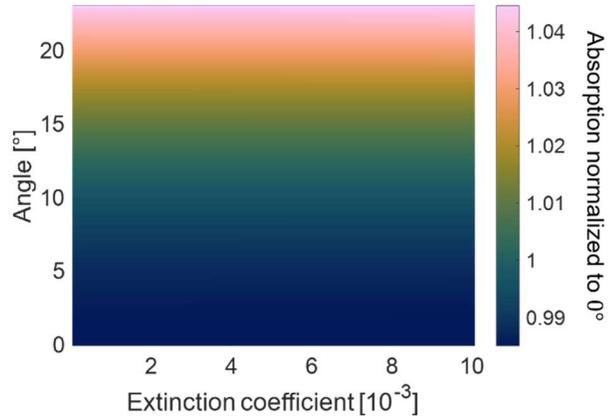

**Fig. S4 Independence of normalized angle-resolved absorption on extinction coefficient.** Absorption of the 50-nm thick silicon nitride membrane normalized by the absorption at normal incidence as a function of extinction coefficient and incidence angle. The angular trend is the same irrespective of extinction coefficient.

The normalized angle-dependent absorption does not depend on the value of the extinction coefficient. This observation turns out to be useful for experimental extraction of the driving power versus angle (Supplementary Notes 1 and 4).

## Extraction of silicon nitride's complex refractive index from ellipsometry

To extract the refractive index of the silicon nitride membrane as shown in Fig. S5, we perform spectroscopic ellipsometry (J.A. Woolam M-2000) measurements on a substrate-supported area of the device. Specifically, the refractive index of the silicon nitride device at 514 nm is determined to be $n = 2.22$ by fitting the spectroscopic data to a Cody-Lorentz oscillator model. We note that while one could technically also obtain the extinction coefficient from this model, it is too small ($\kappa < 10^{-3}$) to be determined reliably with ellipsometry[2].

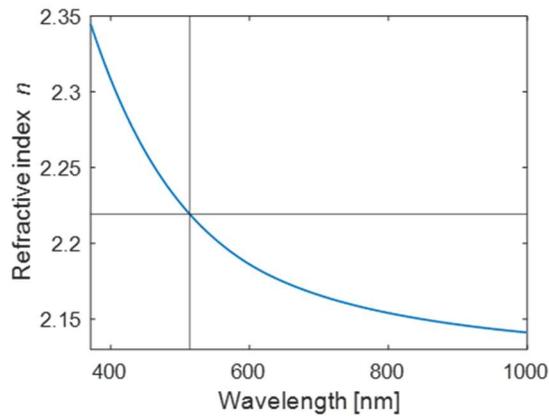

**Fig. S5 Refractive index of silicon nitride from ellipsometry measurements.** Refractive index obtained from fitting to a Cody-Lorentz oscillator model. The extracted value at the driving wavelength of 514 nm is $n = 2.22$.

# Extraction of extinction coefficient based on measured resonance frequency shift

The mechanical resonance frequency shift of the device due to heat generated via optical absorption is used in the main text to extract the driving laser power. Interestingly, by referring to a configuration where the driving power can be precisely determined, the same resonance frequency shift can be used to extract the heat absorption of the membrane. Such a calculation assumes the system reaches thermal equilibrium via heat conduction through the springs of the membrane. This is described by a 1D heat transfer model, while neglecting the effect of thermal emission. The device stiffness is governed by bending of the springs (due to their significant compliance relative to the central pad), with the central pad being the source of heat flux. Following Schmid et al.[3], the heat flux $Q$ through all four springs is equal to the absorbed power,

$$Q = -4\kappa_{\text{th}} A_s \frac{\partial T}{\partial x} = AP, \tag{S13}$$

where $\kappa_{\text{th}}$ is the thermal conductivity, $A_s$ the cross-sectional area of each spring, $\partial T/\partial x$ the temperature gradient along the spring, $A$ the absorption, $P$ the incident power and $x$ is the curvilinear coordinate along the spring. Further assuming a linear gradient along each of the four springs of unfolded length $L$, i.e., $\partial T/\partial x = \Delta T/L$, we can relate the temperature difference between the pad and substrate $\Delta T$ to absorption using

$$\Delta T = \frac{L}{4\kappa_{\text{th}} A_s} AP, \tag{S14}$$

with temperature inducing a thermal strain and thus change in tensile stress. The spatially averaged stress $\langle \Delta \sigma \rangle$ in each spring depends on the spatially averaged temperature change $\langle \Delta T \rangle = \Delta T/2$ as

$$\langle \Delta \sigma \rangle = \alpha_{\text{th}} E \langle \Delta T \rangle = \frac{1}{2} \alpha_{\text{th}} E \Delta T, \tag{S15}$$

where $\alpha_{\text{th}}$ is the coefficient of thermal expansion and $E$ the material's Young's modulus.

The mechanical resonance frequency of the device for zero incident power depends on the tensile stress in the springs, $\sigma$, via $\omega_m \approx \beta\sqrt{1 + \gamma\sigma}$, where $\beta$ is the resonance frequency in the absence of stress and $\gamma$ is the collective mechanical compliance of the device's springs. For nonzero incident power, $P$, the stress $\sigma$ will be reduced by $\langle \Delta \sigma \rangle$, resulting in

$$\omega_m(P) = \beta\sqrt{1 + \gamma(\sigma - \langle \Delta \sigma \rangle)} = \omega_m(0)\sqrt{1 - \frac{\gamma \langle \Delta \sigma \rangle}{1 + \gamma\sigma}}. \tag{S16}$$

Further assuming the stress-dependent term to dominate, i.e., $\gamma\sigma \gg 1$, and Taylor expanding for small changes to the tensile stress, i.e., $\langle \Delta \sigma \rangle \ll \sigma$, we obtain the following expression for the power-dependent resonance frequency

$$\omega_m(P) \approx \omega_m(0)\left[1 - \frac{\langle \Delta \sigma \rangle}{2\sigma}\right] = \omega_m(0)\left[1 - \frac{1}{16}\frac{\alpha_{\text{th}}}{\kappa_{\text{th}}}\frac{E}{\sigma}\frac{L}{A_s} AP\right], \tag{S17}$$

where we have used the relation,

$$\langle \Delta\sigma \rangle = \frac{\alpha_{\text{th}}}{8\kappa_{\text{th}}} \frac{E}{\sigma} \frac{L}{A_s} AP, \tag{S18}$$

with $A_s$ being the effective cross-sectional area of the springs, and $A$ is the membrane's absorption. Solving equation (S17) yields:

$$A = \frac{16\kappa_{\text{th}}\sigma A_s}{\alpha_{\text{th}} E L} \frac{1}{P}\left(1 - \frac{\omega_m(P)}{\omega_m(0)}\right). \tag{S19}$$

The responsivity $\mathcal{R}$ of the mechanical resonance frequency, $f_m$ where $\omega_m = 2\pi f_m$, to the driving power, $P$, defined according to the equation $f_m(P) = f_m(0) - \mathcal{R}P$, is used in the main text to characterize the slope of Fig. 4c (see also equation (S3)). Using this responsivity to simplify equation (S19) gives:

$$A = 32\pi \frac{\kappa_{\text{th}}\sigma A_s \mathcal{R}}{\omega_m(0)\alpha_{\text{th}} E L}. \tag{S20}$$

For device parameters $L = 440$ µm and $A_s = w \times t = 2$ µm $\times$ 50 nm and the following assumed material properties[4,5] $\alpha_{\text{th}} = 2.3 \times 10^{-6}$ K$^{-1}$, $\kappa_{\text{th}} = 2.5$ Wm$^{-1}$K$^{-1}$, $E = 270$ GPa, $\sigma = 100$ MPa, we extract an absorption of $A = 1.4 \times 10^{-4}$ for a responsivity of $\mathcal{R} = -0.9$ Hz/µW (see Fig. 4). We can further calculate the extinction coefficient, $\kappa$, for this extracted absorption by solving equation (S9) for $\kappa$, which yields $\kappa = 1.36 \times 10^{-4}$.

The associated absorption of $A = 1.4 \times 10^{-4}$ is substituted into equation (S14) to obtain a temperature increase of $\Delta T \approx 1.5$ K. For such a small temperature increase, thermal conduction ($\kappa_{th}\Delta T/L$) will dominate over thermal emission ($\sigma\epsilon(T^4 - T_0^4)$), allowing us to neglect the effect of the latter.

## Supplementary Note 4: Calculation of normalized force versus angle of incidence

According to the previous section, the values of the normalized force versus the angle of incidence (Fig. 5b) are calculated according to:

$$\frac{F_z(\theta)}{F_z(0)} = \frac{S_{F-P}(\theta)}{S_{F-P}(0)} = \frac{S_{F-\alpha}(\theta)}{S_{F-\alpha}(0)} \frac{S_{f_0-\alpha}(\theta)}{S_{f_0-\alpha}(0)} \frac{\mathcal{R}(\theta)}{\mathcal{R}(0)}, \tag{S21}$$

where $S_{F-\alpha}(\theta)$ and $S_{f_0-\alpha}(\theta)$ are obtained from linear fits according to equations (S5) and (S6), as shown in Fig. 4a and Fig. 4c. The responsivity $\mathcal{R}(\theta)$ is linearly proportional to the optical absorption $A$. Therefore $\mathcal{R}(\theta)/\mathcal{R}(0) \propto A(\theta)/A(0)$. The absorption generally depends on the optical extinction coefficient, whose precise determination can be challenging. However, for membranes, thin-film interference calculation reveals that the ratio $A(\theta)/A(0)$ does not depend on the extinction coefficient, but only on the refractive index of the layer and its thickness (see Supplementary Note 3, Fig. S4). Specifically, for our membrane thickness of 50 nm and refractive index of the silicon nitride $n = 2.22$, the ratio changes by less the 6% in the range from 0° to 23°, i.e., $A(23°)/A(0°) = 1.06$. By accounting also for this absorption trend, we ensure precise force characterization.

## Supplementary Note 5: Error estimation

The error bars in Fig. 5 are calculated according to a confidence interval of one standard deviation (68%) by Gaussian error propagation for uncorrelated variables. Specifically, the force error used to calculate the error bars in Fig. 5b are obtained by propagating the uncertainties of $S_{F-\alpha}(\theta)$ and $S_{f_0-\alpha}(\theta)$:

$$\Delta F_z(\theta) = F_z(\theta) \sqrt{\left(\frac{\Delta S_{F-\alpha}(\theta)}{S_{F-\alpha}(\theta)}\right)^2 + \left(\frac{\Delta S_{f_0-\alpha}(\theta)}{S_{f_0-\alpha}(\theta)}\right)^2}. \tag{S22}$$

The uncertainties $\Delta S_{F-\alpha/f_0-\alpha}(\theta)$ are calculated while accounting for both statistical error $\Delta S^{\text{sta}}_{F-\alpha/f_0-\alpha}(\theta)$ and instrumental error $\Delta S^{\text{ins}}_{F-\alpha/f_0-\alpha}(\theta)$:

$$\Delta S_{F-\alpha/f_0-\alpha}(\theta) = \sqrt{\left(\Delta S^{\text{sta}}_{F-\alpha/f_0-\alpha}(\theta)\right)^2 + \left(\Delta S^{\text{ins}}_{F-\alpha/f_0-\alpha}(\theta)\right)^2}. \tag{S23}$$

Statistical errors are obtained from the standard errors of the linear fits according to equations (S5) and (S6), while taking three distinct time series measurements for each of the six different powers. See Methods for full characterization of the time series parameters. The instrumental error of the force slope $S_{F-\alpha}(\theta)$ is governed by the temperature-dependent multiplication factor of the APD due to temperature fluctuations in the laboratory, and is assumed to be 3% of the corresponding value, according to the manufacturer specifications. The statistical error of the power slope is taken to be $\Delta S^{\text{ins}}_P(\theta) = 0$, because the APD readout error insignificantly affects the resonance frequency determination. Particularly, this holds true for error variations on a time scale much smaller than the mechanical resonance cycle $1/f_0$.

Errors in the reflection measurements are also calculated by uncertainty propagation of the corresponding errors, in equation (S12), for uncorrelated random variables. The errors due to fluctuations of the temporally resolved reflected powers $R^{\text{exp}}_{\text{sam}}$ and $R^{\text{exp}}_{\text{sub}}$ obtained from short-time Fourier transform (see Supplementary Note 3) are calculated according to a confidence interval of one standard deviation (68%). The error in $R^{\text{theo}}_{\text{sub}}$ is calculated according to uncertainty of 1 nm in the 100-nm silicon nitride layer thickness on top of the silicon chip.

## Supplementary Note 6: Calculation of radiation pressure forces from COMSOL simulation

Evaluating the Maxwell stress tensor over a closed circular path around the illuminated 1D pad of thickness $t = 50$ nm and width $l = 40$ μm in COMSOL Multiphysics results in optically induced pressures per unit depth. Because the oblique incidence angles are modelled by rotating the pad by an angle $\theta$, this pressure needs to be transformed to the frame of the rotated pad. Finally, the radiation pressure force $F_z$ per unit depth can be calculated from the pressure $p_z$ in the pad frame according to

$$F_z(\theta, x) = \int_{-\cos(\theta)l/2}^{\cos(\theta)l/2} p_z(\theta) I(r)\, dr, \tag{S24}$$

where $I(r)$ is the 1D Gaussian intensity distribution of the beam

$$I(r,x) = \frac{P_0}{w\sqrt{\pi/2}} e^{-2(x+r)^2/w^2}, \tag{S25}$$

with beam width $w$ and offset $x$ relative to the center of the pad corresponding to the origin.

Note that the integral of the Gaussian beam intensity over the pad width accounts for potential dependence of the incident power on angle $\theta$. Analytically evaluating equation (S24) yields:

$$F_z(\theta, x) = \frac{P_0}{2} p_z(\theta) \left[ \text{erf}\left(\frac{2x + l\cos(\theta)}{w}\right) - \text{erf}\left(\frac{\sqrt{2}(x - l\cos(\theta)/2)}{w}\right) \right]. \tag{S26}$$

# Supplementary Note 7: Lightsail devices to verify the effect of edge scattering on the radiation pressure force

To experimentally show the effect of edge scattering on the optical force, we fabricate additional devices with spring-supported squared pads of three side lengths: 40 μm, 60 μm and 80 μm (Fig. S6). In contrast to the device presented in the main text, these devices have thickness of 100 nm. The awl-shaped springs are designed such that the fundamental mode frequencies of the devices are close to 8.4 kHz by running prestressed eigenfrequency studies in COMSOL Multiphysics.

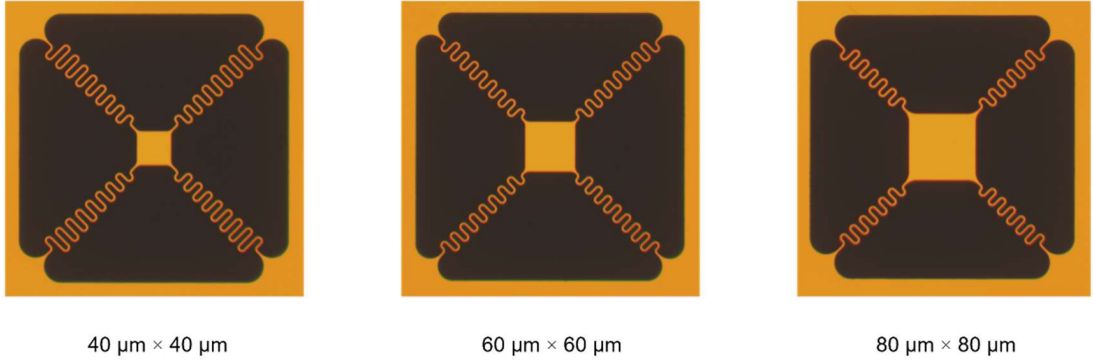

40 μm × 40 μm          60 μm × 60 μm          80 μm × 80 μm

**Fig. S6 Devices for characterization of optical force reduction due to edge scattering.** Microscope images of 100-nm thick, square silicon nitride lightsail membranes suspended by four awl-shaped serpentine springs with varying side length of 40 μm, 60 μm and 80 μm (from left to right).